# Explaining the Unexplainable: A Systematic Review of Explainable AI in Finance

Md Talha Mohsin[a*], Nabid Bin Nasim[b]

[a] Collins College of Business, University of Tulsa

[b] Faculty of Biological Sciences, University of Dhaka

Corresponding Author: Md Talha Mohsin; mdtalhamohsin@utulsa.edu

**Abstract:** Practitioners and researchers trying to strike a balance between accuracy and transparency center Explainable Artificial Intelligence (XAI) at the junction of finance. This paper offers a thorough overview of the changing scene of XAI applications in finance together with domain-specific implementations, methodological developments, and trend mapping of research. Using bibliometric and content analysis, we find topic clusters, significant research, and most often used explainability strategies used in financial industries. Our results show a substantial dependence on post-hoc interpretability techniques; attention mechanisms, feature importance analysis and SHAP are the most often used techniques among them. This review stresses the need of multidisciplinary approaches combining financial knowledge with improved explainability paradigms and exposes important shortcomings in present XAI systems.

## 1. Introduction

The widespread adoption of artificial intelligence (AI) has caused people to reassess their own positions (Cao et al. 2024). Thanks to significant developments in processor capacity and advancements in optimization methods, we are seeing widespread acceptance of automated decision making by AI (Sachan et al. 2020). Because of these AI system's popularity, they have also received significant attention in recent popular press (Bornstein 2016) (Harford 2014) (Hawkins 2017) (Kuang 2017) (Pavlus 2017) (Mikulak 2017). Through the use of modern computational approaches, decision-making processes in a variety of financial industries have also been revolutionized.

In the field of finance, AI has surprised everyone (Hidayat, Defitri, and Hilman 2024). Both the large amount of data that is available and the increasing reliance on machine learning (ML) models have impacted the world of finance substantially. Specially the growing outflow of data created by consumers, investors, businesses, and governments have helped AI transform the financial sector (Bahoo et al. 2024). Also the professionals in the finance sector are growingly fascinated in "alternative data" outside the scope of macroeconomic indicators, securities pricing, and basic business knowledge which include satellite imagery, news stories, phone records, and social media post, which require more understanding and have AI implications (Goodell et al. 2023). AI is not only a phase; rather, it is a presence that is fundamentally changing how companies are handling their assets and finances.

Despite their remarkable success in producing high-performance models, many of these AI and ML techniques have come under fire for their lack of transparency and interpretability. In many scenarios, machine learning models acquire great accuracy, usually at the price of poor explainability (Giudici and Raffinetti 2023). The more complicated these models get, the more often their decision-making processes remain opaque, which creates a barrier for stakeholders trying to know why and how a particular choice was taken. For applications where sophisticated ML models are part of how stakeholders come to make decisions, giving user-centered explanations is very important (Zhou et al. 2023). Especially in the financial sector, where decision-makers including investors, authorities, and financial analysts depend on not only accurate data but also the justification for these to guarantee informed, responsible, and accountable actions.

Reacting to this problem of interpretability, explainable artificial intelligence (XAI) has become a major focus of study. Although expert Systems researchers have used explanation approaches before (Clancey 2014) (Clancey 1986) (McKeown and Swartout 1987) (Moore and Swartout n.d.), the term explainable AI (XAI) was coined by DARPA (Gunning and Aha 2019). Explainability can be thought of as an active feature of a model, referring to any action or process that a model takes to elucidate its internal operations (Arrieta et al. 2020). The quality of "being an explanation" is an interaction rather than a characteristic of assertions (Hoffman et al. 2019). Explanation techniques have advanced significantly as XAI is now acknowledged as a must rather than merely an option (Wazid et al. 2022). What the user requires, what information they already possess, and most importantly, their objectives, determines what constitutes an explanation. This raises the question of why a particular user needs an explanation, which brings up the function and context of the AI system (software, algorithm, tool). By integrating interpretability restrictions into the model's structure, XAI approaches make the models easier to understand (Rai 2020).

Although scholars have paid more and more attention to the adoption of AI technologies in a wide spectrum of financial applications in recent years, the existing literature is rather broad and heterogeneous in terms of research questions, level of analysis and method, making it difficult to draw solid conclusions and to understand which research areas require further investigation (Bahoo et al. 2024). Despite the fact that a large number of machine learning interpretability research and approaches have been established in academia, they hardly ever make up a significant portion of machine learning pipelines and workflows (Linardatos, Papastefanopoulos, and Kotsiantis 2020). As such, focusing on its theoretical foundations, approaches, and applications across many financial sectors, this review article attempts to give a thorough understanding of the present situation of XAI in finance. By combining current research, this paper highlights the most significant XAI approaches and their contributions to improve model transparency and interpretability.

For our selected papers, we applied co-authorship networks, trend analysis, and keyword frequency analysis to understand the current scenarios. Then, based on their significance and influence in the field, we divided the chosen papers into seven streams: risk management, fraud detection, time series forecasting, financial analysis and decision-making, credit evaluation and scoring, financial modeling and prediction, and trading and investment. We conducted citation analysis, and co-citation analysis among several bibliometric approaches to understand the current situation. Our evaluation aims to underline the advantages and drawbacks of present XAI approaches as well as the limitations to increase AI system justice, responsibility, and openness.

The remainder of this study is structured as follows: Section 2 discusses background and context, Section 3 outlines the data and methodology, Section 4 presents the findings and discussion, and Section 5 concludes the study.

## 2. Related Works

Artificial Intelligence (AI) has gained significant traction in recent years and, with the right application, might meet or surpass expectations in a wide range of application areas (Arrieta et al. 2020). However by today's standards, it is not advisable to blindly trust the output of AI because of the significant influence of adversarial instances, trustability, and data bias in machine learning (Das and Rad 2020). Even when we understand the mathematical underpinnings of machine learning (ML) architectures, it is frequently impossible to get insight into how the models operate internally (Goebel et al. 2018); explicit modeling and reasoning techniques are required to clarify how and why a particular outcome was obtained. That's where Explainable Artificial Intelligence (XAI) comes in. Explainability is the ability of an interested stakeholder to understand the major reasons of a model-driven decision (Bussmann et al. 2021). XAI is a component of the third wave of AI (Adadi and Berrada 2018), a new generation of AI technologies whose goal is to create algorithms that can accurately explain themselves. It entails the capacity of an artificial intelligence system to provide coherent and easily accessible explanation for decisions and actions it makes (Talaat et al. 2024). XAI makes the end user, who relies on decisions, suggestions, or actions made by an AI system, the focus, as that person must comprehend the reasoning behind the system (Gunning and Aha 2019).

A machine learning model's explainability and prediction accuracy are typically inversely correlated (Xu et al. 2019); the better the prediction accuracy, the lower the model's explainability. XAI aims to create approaches that, without compromising their prediction accuracy, increase the transparency and interpretability of artificial intelligence and machine learning models. Clear descriptions of model outputs help XAI practitioners and users to understand the underlying processes and justification for AI-driven judgments. explainability can be viewed as an active characteristic of a model, denoting any action or procedure taken by a model with the intent of clarifying or detailing its internal functions. Understandability stands out as the most crucial idea in XAI (Arrieta et al., 2020), which is closely related to both interpretability and transparency: interpretability gauges how well a human can comprehend a model's choice, whereas transparency describes a model's ability to be intelligible by a human on its own. Understandability and comprehensibility are related in that both depend on the audience's capacity to comprehend the information presented in the model.

In order to cover every facet of XAI, concentrate on providing answers to the questions "What, Why, and How" (Arrieta et al. 2020), (Das and Rad 2020), (Khaleghi n.d.). The "What" aims to clarify the current meanings of explainable AI and the significance of elucidating the function of a user. The "Why" gives a summary of the main objectives of XAI research, such as establishing credibility, fulfilling legal obligations, detecting bias, guaranteeing generalization in AI models, and debugging. The "How" section looks at how to achieve explainability prior to the modeling phase, including how to fully comprehend and record the datasets used in modeling. "How" section examines the methods for attaining explainability before the modeling stage, including techniques for thoroughly understanding and documenting the datasets utilized in modeling. XAI seeks to explain how AI models make decisions in a way that people can grasp, therefore increasing their transparency.

XAI comprises a wide range of strategies and methodologies that can be broadly classified into two types: post-hoc and ante-hoc explainability. Post-hoc explainability refers to methods used after a model has been trained to provide reasons for its predictions, whereas Ante-hoc explainability entails creating intrinsically interpretable models which are transparent by definition and produce interpretable outcomes directly (Fig 1).

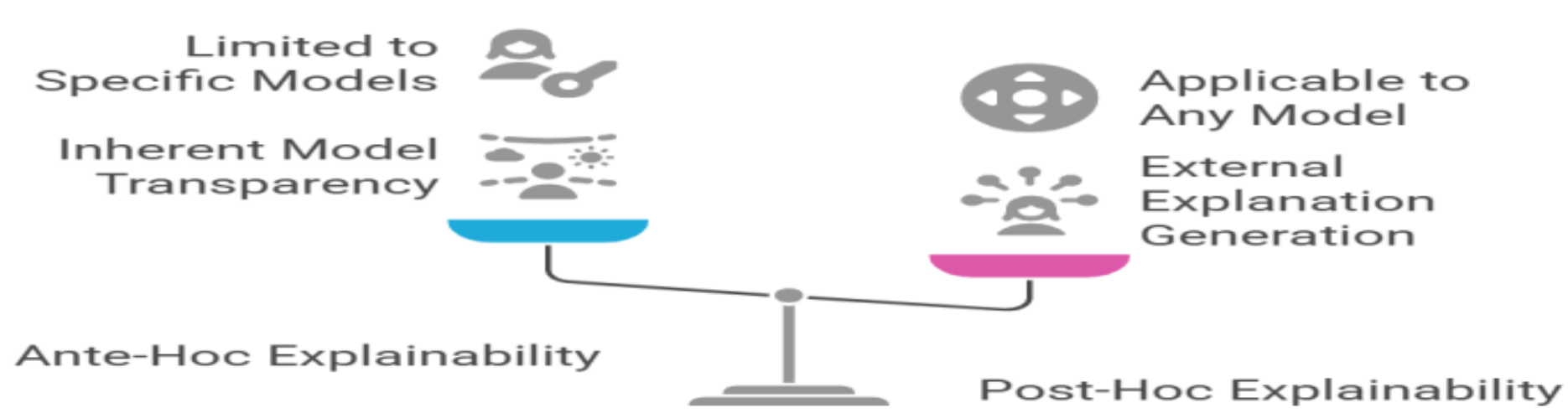

**Fig 1.** Ante-Hoc vs. Post-Hoc Explainability

Also, there are two types of machine learning models based on their understandability: white-box models and black-box models. These models are a way to group AI models that affect the need for XAI methods. Then, XAI techniques are used to Fig out what these models mean. The difference between white-box and black-box models is directly linked to the difference between post-hoc and ante-hoc explainability. Ante-hoc explainability fits well with white-box models because these models are meant to be interpretable and give information about how they make decisions without needing to be interpreted further. These are better when trust and openness are very important. On the one hand, there are "black-box" models, such as deep learning (LeCun, Bengio, and Hinton 2015) and ensembles (Chen and Guestrin 2016) (Liaw 2002) (Polikar 2012). Specially those based on deep learning, frequently thought of as "black-box" models (Lakkaraju et al. 2017) (Das and Rad 2020) (Gunning and Aha 2019) (Dwivedi et al. 2023). These black-box models need post-hoc explainability tools to help us understand how they make decisions when they aren't being clear. (Fig 2) depicts the decision-making process when selecting the appropriate model type for XAI.

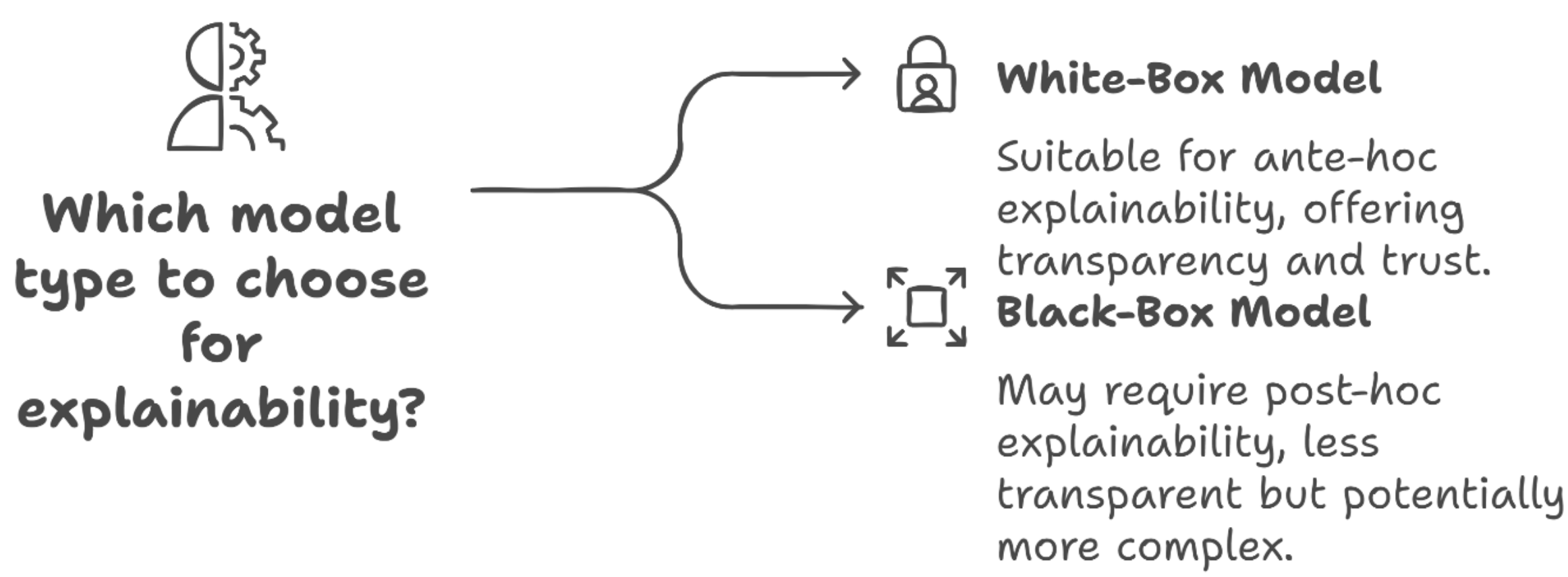

**Fig 2.** Choosing Between White-Box and Black-Box Models for Explainability

XAI helps release the power of machine learning algorithms to challenge the black-box character (Nallakaruppan, Balusamy, et al. 2024). White box models are transparent and very simple to comprehend, while black box models are opaque (Dwivedi et al. 2023). Since they enable improved quality assurance of black box ML models, XAI tools are a significant addition to the data science toolkit, specially to understand the characteristics of the data collection and domain expertise helps them to complement other facets of quality assurance, including several approaches of model performance testing (Bracke et al. 2019).

Although modern machine learning models are more easily available than ever, it has been difficult to design and implement systems that support real-world financial applications (Nallakaruppan, Chaturvedi, et al. 2024). This is mostly due to their lack of transparency and explainability— both of which are fundamental for establishing trustworthy technology. In the financial domain, justification for the choice of action is just as important as the result itself. XAI's ability to advance justice and lower bias in artificial intelligence systems is among its most significant advantages for the financial industry. For example, XAI techniques can help find the factors underlying the discrimination if a credit score model is unfairly excluding particular demographic groups thereby enabling corrective action. Maintaining ethical AI principles depends on this, especially in lending where biassed models might cause financial exclusion. By giving financial analysts, investors, and officials better knowledge of how artificial intelligence models operate, XAI also helps with decision-making. XAI technology can find economic indicators, market data, or historical trends driving model forecasts whether used for stock price prediction or portfolio management. This not only boosts model confidence but also lets financial decision-makers act with certainty based on these realizations. XAI can assist in stock price forecasting, for instance, by dissecting the pertinent elements—such as earnings reports, market sentiment, or macroeconomic data—so enabling an investor to understand why a given model forecasts an increase in stock price.

Apart from the complexity of financial models, the demand for XAI in finance results from rising expectations for openness from authorities and the public overall. Financial institutions—including banks, asset managers, and insurance firms—are expected more and more to justify their AI-driven decisions to stakeholders. Explainable artificial intelligence models—which include specifics or justifications to make the operation of artificial intelligence clear or simple—are essential.

# 3. Methodology

### 3.1 Data Collection

We systematically searched the Scopus database using a well chosen set of keywords combining finance and Explainable AI (XAI) to identify the relevant papers. We then carefully examined their titles, abstracts, and keywords so that they matched our screening criteria. To ensure the inclusion of recent developments in XAI and its applications in finance, we limited the search for this study to English-language publications published between 2015 and the present.

### 3.2. Identification of Keywords

The keyword selection procedure for this study was intended to capture the most relevant material on the intersection of Explainable AI (XAI) and finance while minimizing irrelevant results **(Table 1)**. The keyword identification involved a thorough investigation of AI and machine learning (ML) glossaries, dictionaries, and academic literature to discover fundamental terminologies used in the area. XAI-related research regularly uses terminology like "explainable AI," "interpretable machine learning," "post-hoc explanations," "transparency," and "model interpretability." Key terms pertinent to the financial domain were also selected, such as "finance," "financial systems," "fraud detection," "credit scoring," "portfolio optimization," and "stock price prediction." This methodical methodology enabled a targeted retrieval procedure to get papers relevant to the research topic.

**Table 1:** Query Terms

| Domain | Keywords |
|---|---|
| XAI | "explainable AI" OR XAI OR "interpretable machine learning" OR interpretability OR "explainable artificial intelligence" OR "model interpretability" OR "post hoc explanations" OR "feature attribution" OR transparency OR explainability OR "Attention Mechanisms" OR "Attention Mechanism" OR SHAP OR LIME OR "Integrated Gradients" OR "Counterfactual Explanations" OR "Saliency Maps" OR "Feature Importance" OR "Partial Dependence Plots" OR "Individual Conditional Expectation" OR "Explainable Neural Networks" |
| Finance | finance OR stock OR "stock price prediction" OR "stock price forecasting" OR "fraud detection" OR "credit scoring" OR loan OR "risk management" OR "risk assessment" OR "Credit risk assessment" OR "portfolio optimization" OR "portfolio management" OR "algorithmic trading" OR "wealth management" OR "financial forecasting" OR ESG OR "investment analysis" OR "personal finance" OR cryptocurrency OR "financial systems" OR "financial technology" OR fintech OR "Regulatory Compliance" OR "Sentiment Analysis for Financial Markets" OR "financial markets" OR "banking" OR "insurance" OR "asset management" OR "derivatives pricing" |

### 3.3. Refinement and Search Optimization

The initial keyword search returned 6,086 publications. Given the large volume of results, a step-by-step screening process was used to identify the most relevant studies. First, we applied scope exclusion criteria to remove papers outside the study’s focus. Specifically, we excluded works that addressed finance without XAI, XAI without financial applications, or AI-in-finance without explainability, resulting in 551 remaining papers.

To maintain quality and credibility, we then applied an impact filter, retaining only papers with at least 50 Scopus citations, which reduced the set to 323 studies. Next, we limited the selection to English-language papers published in peer-reviewed journals or conference proceedings, explicitly excluding preprints, duplicates, and non-peer-reviewed materials. This yielded 237 unique papers.

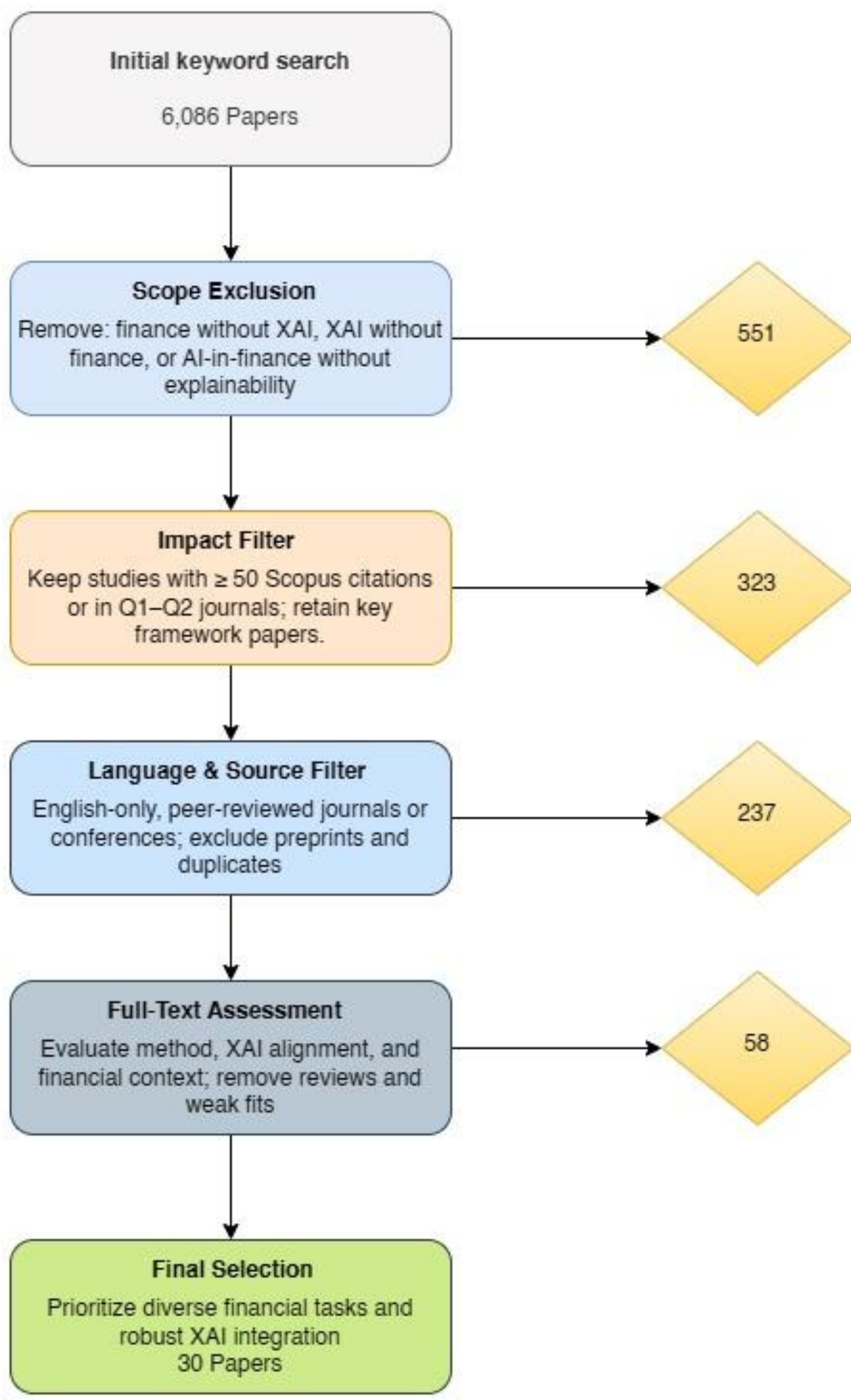

**Fig 3.** Flowchart of Literature Selection and Refinement Process

Each of these papers underwent full-text assessment to evaluate methodological rigor, relevance to financial decision-making, and alignment with XAI frameworks. Review papers and studies only loosely connected to the topic were removed, leaving 58 primary studies. Finally, we selected the most influential and well-aligned papers, resulting in a curated set of 30 studies that met all inclusion criteria and best reflected the objectives of this research (Fig. 3).

# 4. Discussion

### 4.1 General Trends

Fig 4 illustrates the keyword co-occurrence network of the principal research topics and their interrelations to these papers.

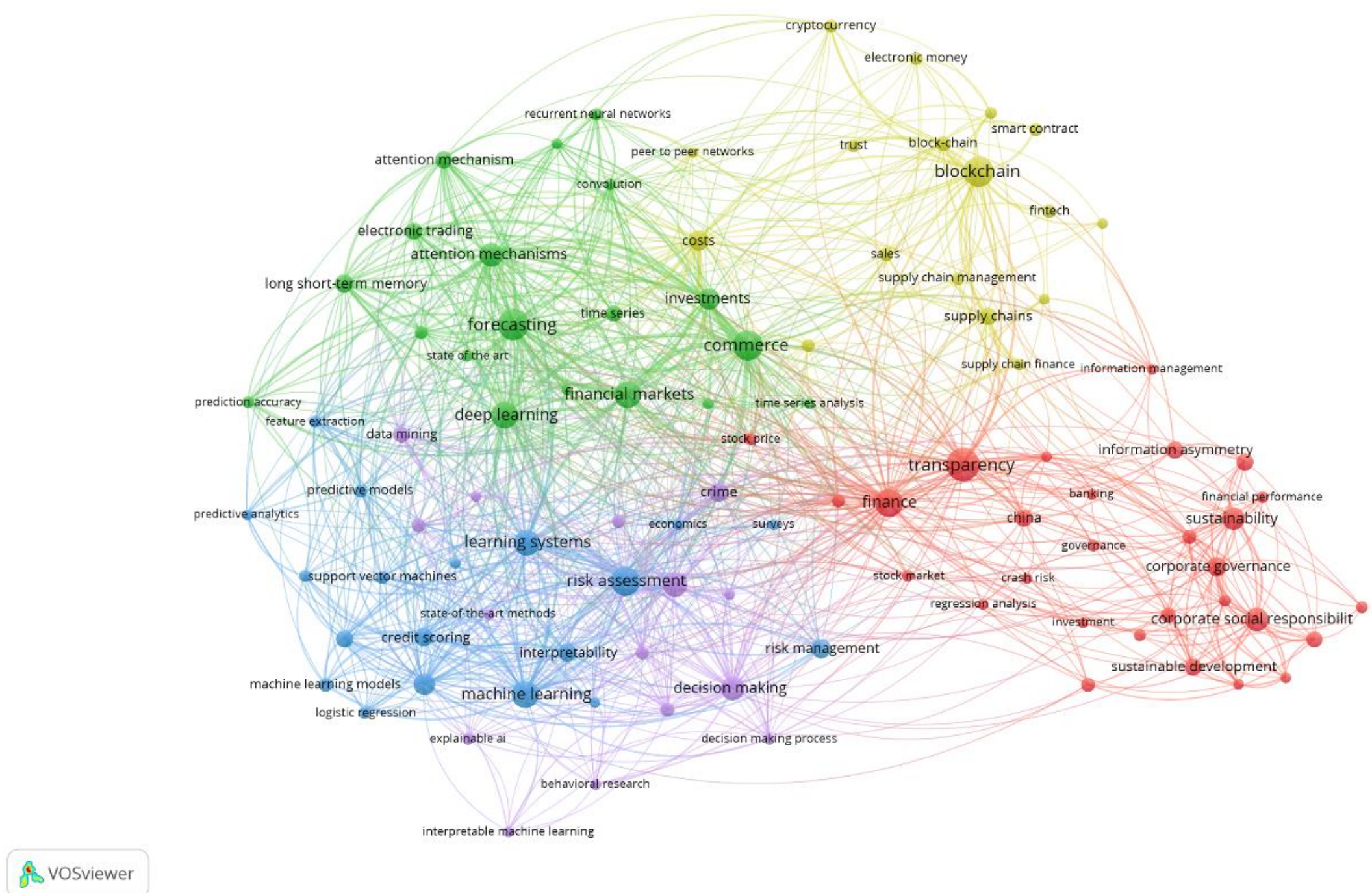

**Fig 4.** Keyword Co-occurrence Network

This network emphasizes the most prevalent keywords, showcasing prominent subfields and emerging trends in literature. Each of the keywords is represented by a circle and the size of the circle denotes the weight of the keyword. The network is segmented into different clusters, underscoring the thematic domains in the literature.

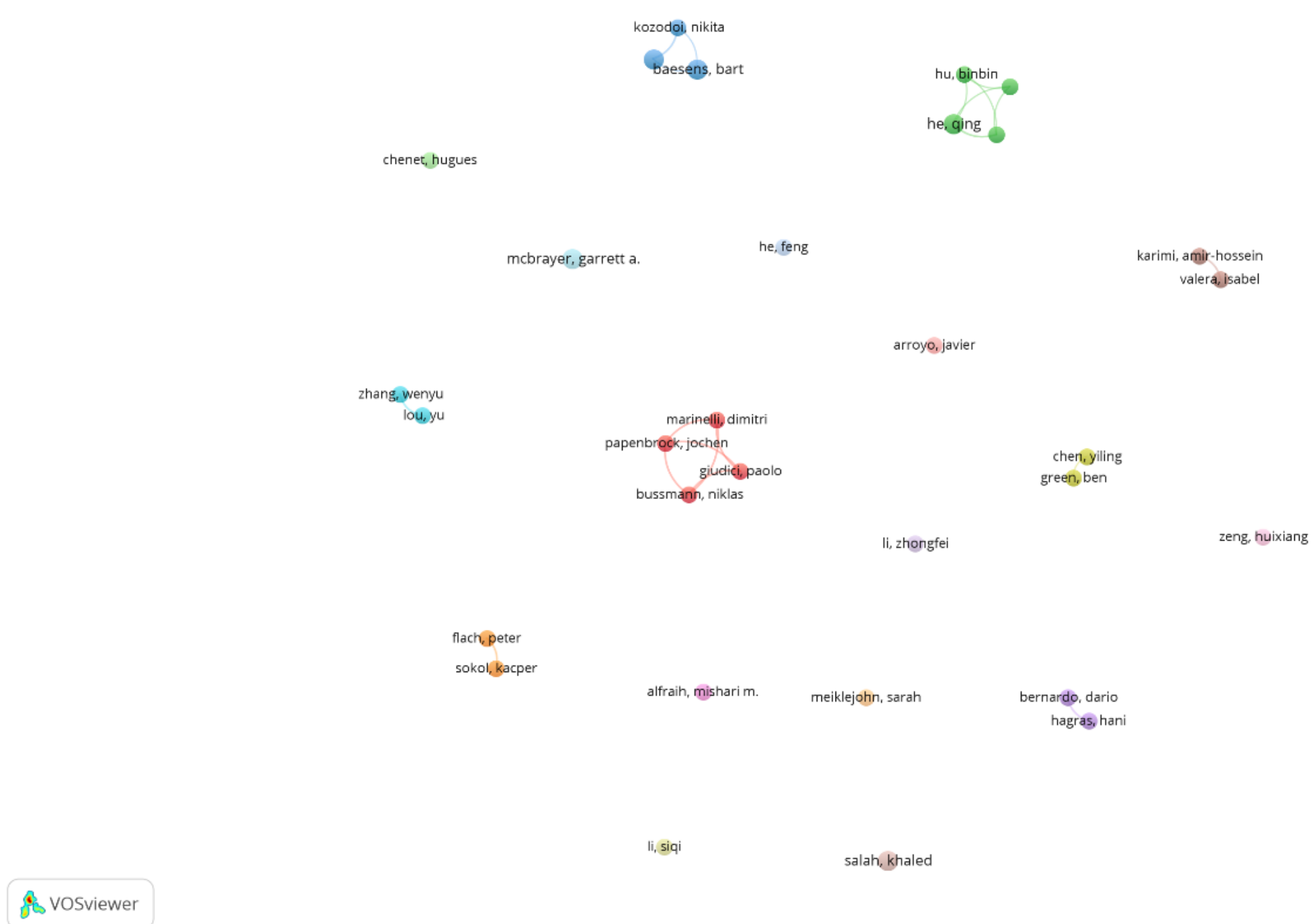

**Fig 5.** Co-authorship Network

(Fig 5) denotes the co-authorship network, visualizing a structural overview of collaboration trends in the field of Explainable AI (XAI) in finance, demonstrating the extent to which researchers collaborate on similar subjects. The network is composed of comparatively small, unrelated clusters, which suggest that research in this field is still relatively niche and dispersed.

### 4.2 Citation Analysis

After further refinement through methodological contribution, journal repute, and citation numbers, the collection was reduced to 30 papers, those judged most pertinent and influential for our research. We split them into seven groups of the financial domain: risk management, fraud detection, time series forecasting, financial analysis and decision-making, credit evaluation and scoring, financial modeling and prediction, and trading and investment **(Table 2)**.

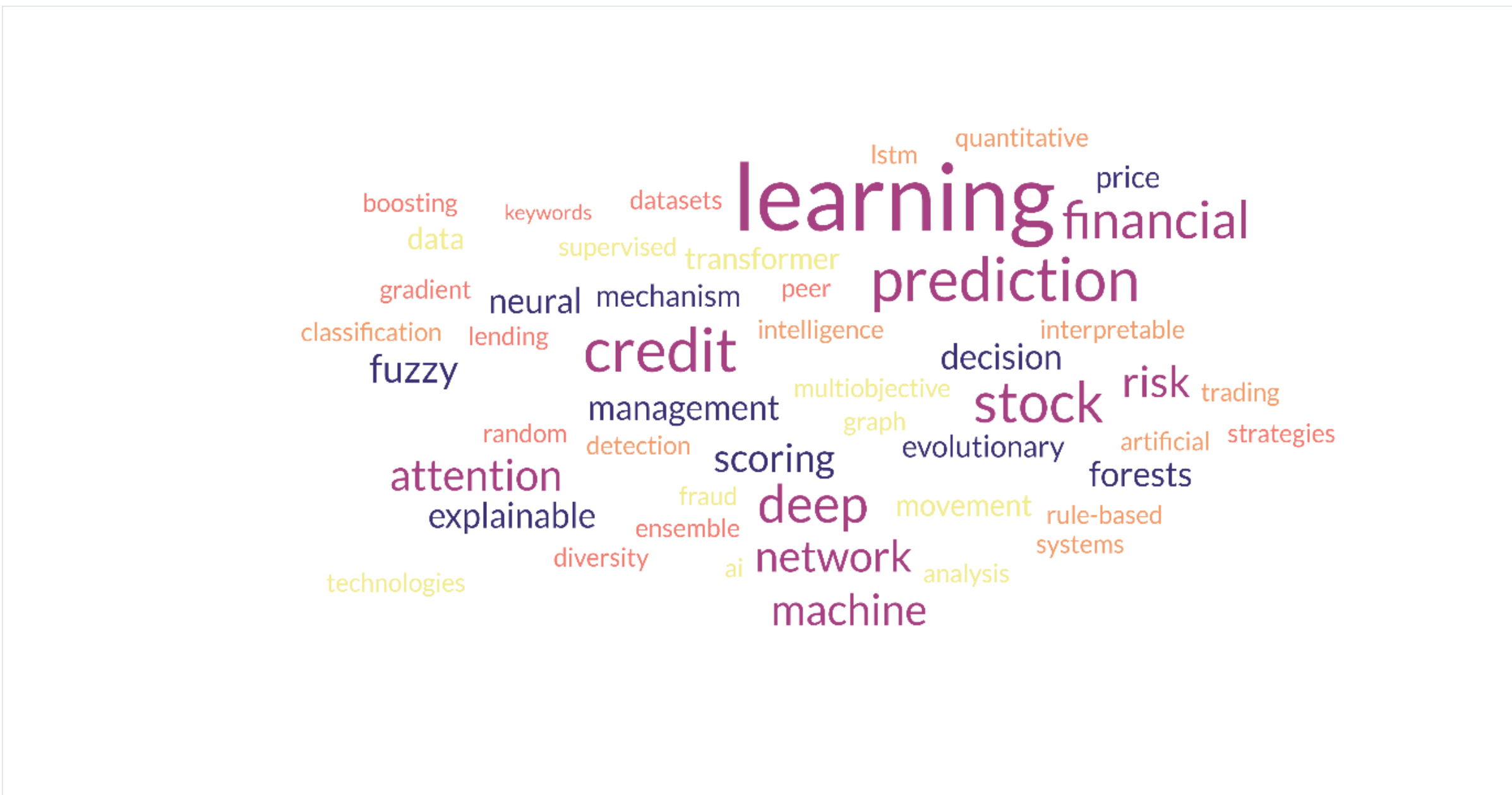

**Fig 6.** Keyword Analysis

(Fig 6) shows the keyword analysis, which offers substantial insights into the prevailing issues and research focal areas at the intersection of finance and Explainable AI (XAI). The Fig highlights the critical aspects of learning, explainability, prediction, credit, stock, finance, and risk management. The increasing emphasis on interpretability, particularly in complex financial models, underscores the importance of fuzzy logic, explainability, and attention mechanisms.

The bar chart in (Fig 7) depicts the distribution of papers published each year. The analysis of publication trends indicates that there has been a lot more research on

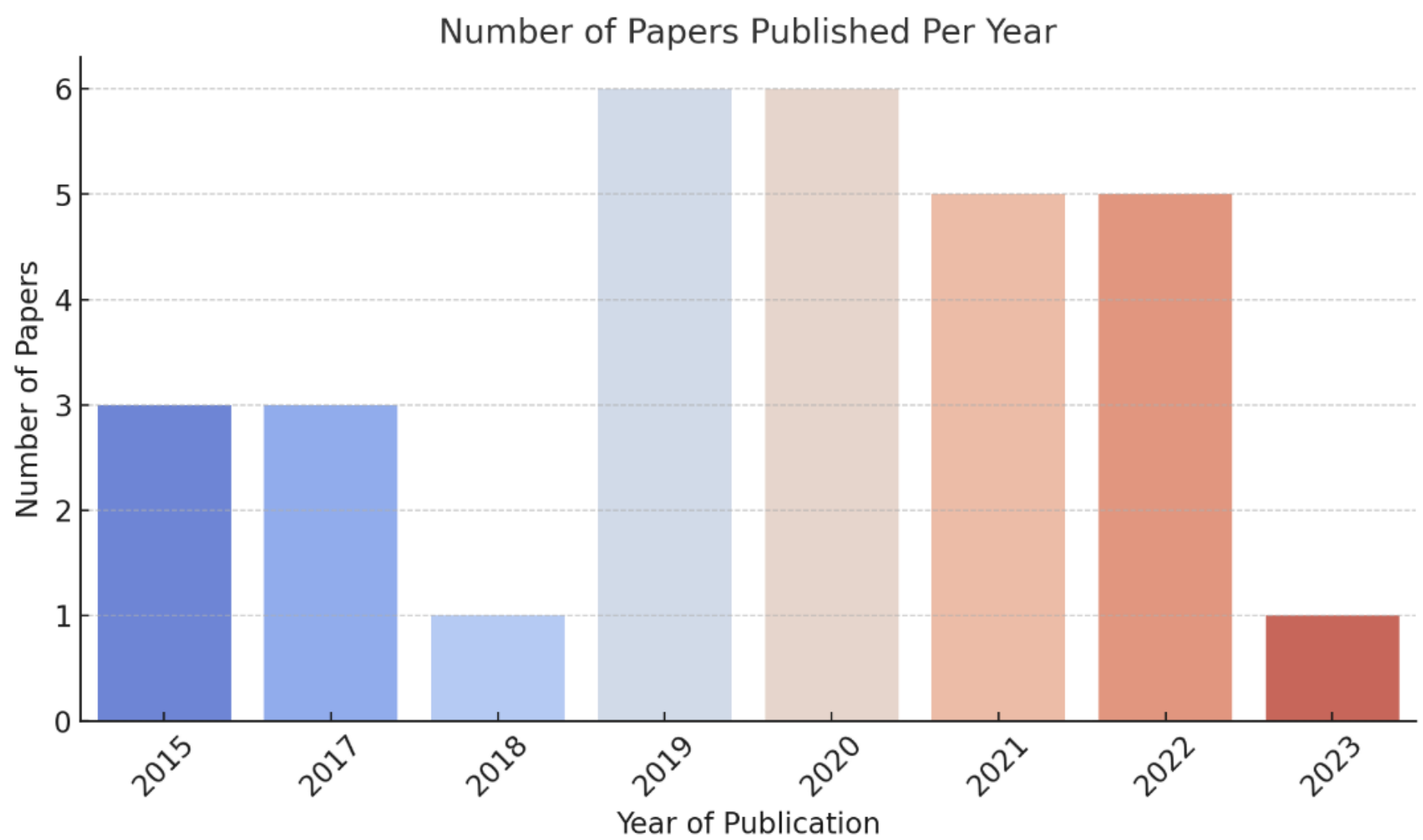

**Fig 7.** Annual Trend of Published Papers

Explainable AI (XAI) in finance after 2019. The interest rate peak in 2019-2020 and remain high in the years following.

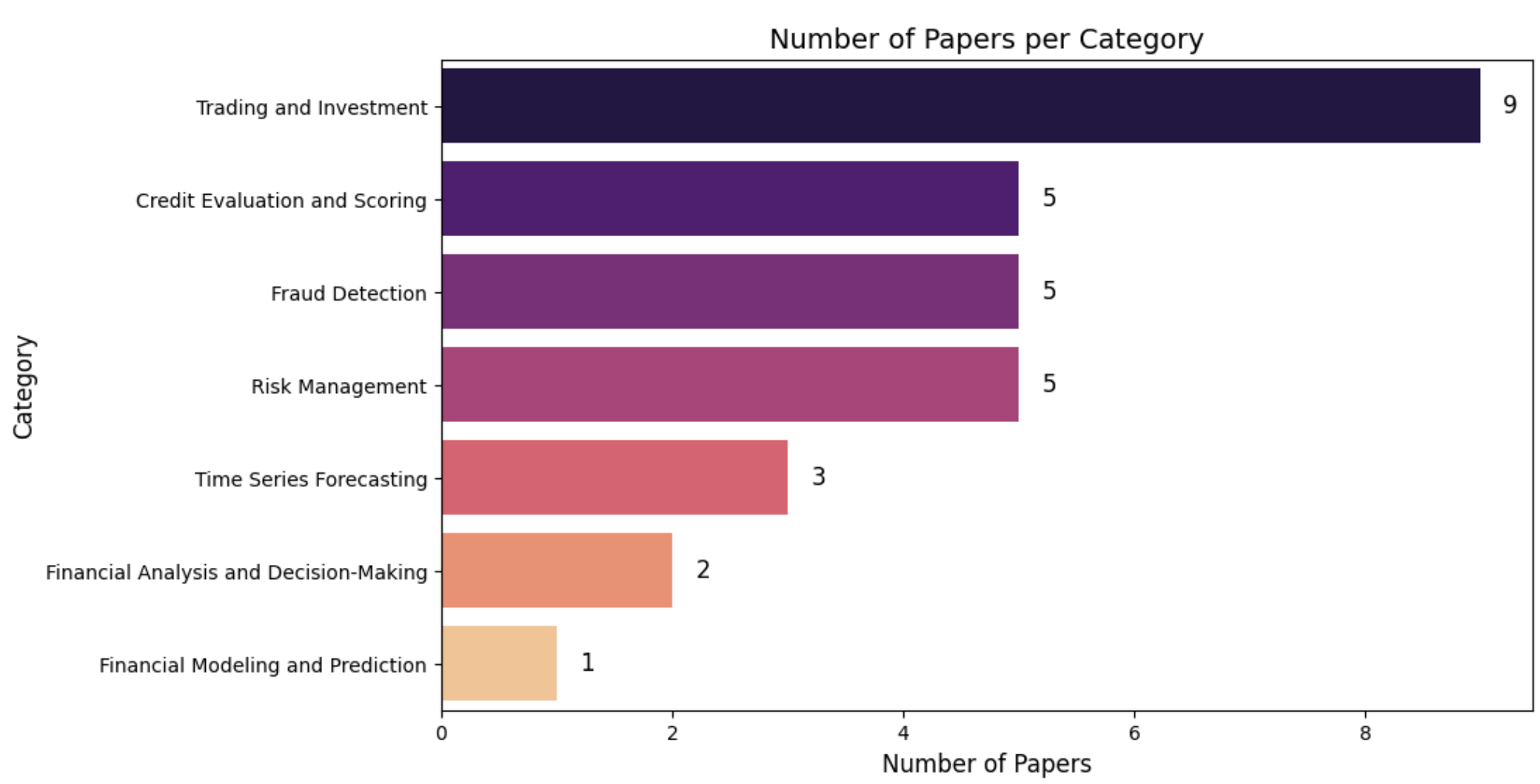

**Fig 8.** Distribution of Research Papers Across Financial Domains

The expansion of research across diverse areas of finance indicates that a lot of attention is being paid to in critical areas. Trading and Investment lead the category, followed by Credit Evaluation, Fraud Detection, and Risk Management (Fig 8).

The leading journal in the distribution of research publications is Expert Systems with Applications, followed by IEEE Transactions on Fuzzy Systems and Decision Support Systems. Top conferences such IJCAI, KDD, and WWW highlight the increasing scholarly and commercial attention on XAI-driven financial applications (Fig 9).

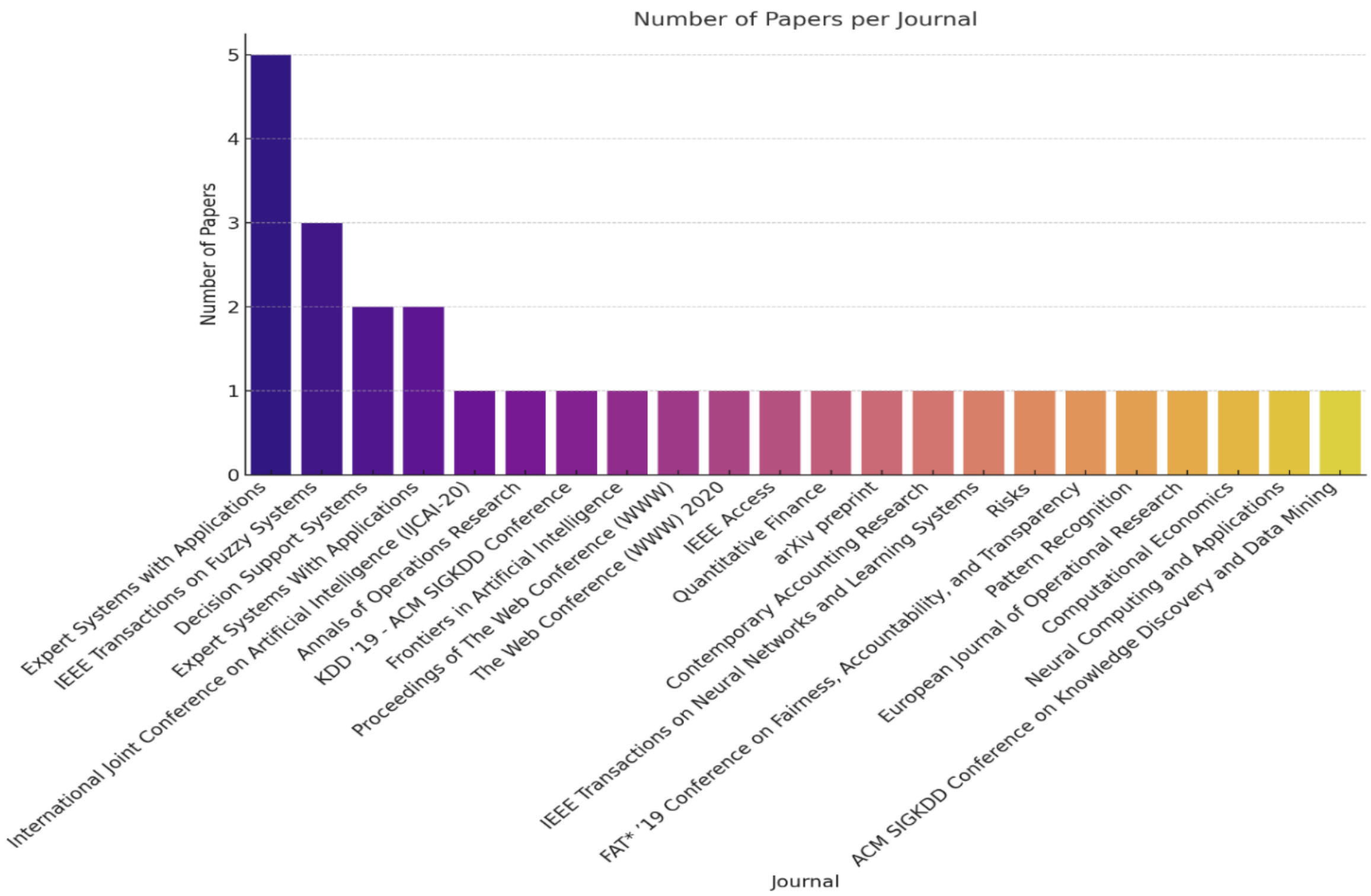

**Fig 9.** Journal and Conference Distribution

As in the case of the usage of XAI methods, attention mechanisms, SHAP, and feature importance analysis emerge as the most frequently employed techniques. The trend also shows the acceptance of XAI approaches is the increasing dependence on attention-based models and SHAP explanations (Fig 10). The adoption of feature attribution techniques such SHAP and LIME has acquired momentum in credit scoring and fraud detection. Furthermore, complementing sophisticated black-box models in high-frequency trading and algorithmic decision-making are post-hoc explainability methods.

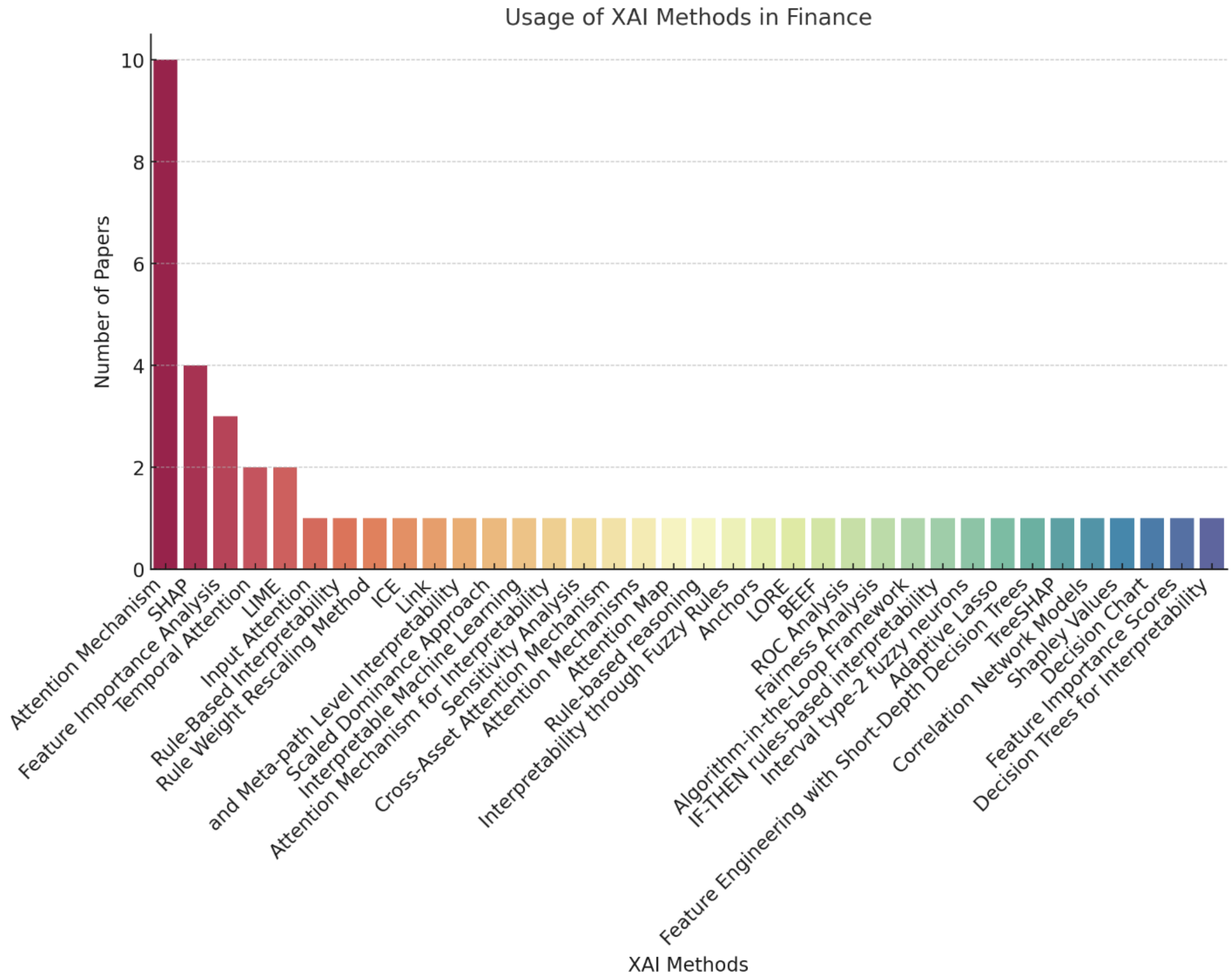

**Fig 10.** Adoption of XAI Techniques

The co-occurrence analysis of XAI and ML algorithms in financial research demonstrates distinct preferences for explainability techniques based on machine learning model complexity (Fig 11). SHAP, attention mechanisms, and feature importance analysis emerge as the most commonly used algorithms, especially when combined with XGBoost, LSTM, CNNs, and Transformers. A notable trend is the strong correlation between post-hoc interpretability techniques, such as SHAP and LIME, and tree-based models like XGBoost and Random Forest, whereas attention mechanisms and temporal interpretability techniques are more commonly used in conjunction with deep learning models, such as LSTMs and Transformers. The heatmap also shows a shift toward hybrid interpretability approaches, which combine different XAI techniques to improve model transparency. The inclusion of decision charts, correlation network models, and sensitivity analysis alongside feature attribution methods indicates a desire to create a multi-layered interpretability framework that balances local and global explanations. The findings show a divergent trajectory in explainability strategies, with traditional financial models relying on feature attribution and rule-based reasoning, and deep learning-driven financial systems requiring layer-wise attention mechanisms and neural network transparency tools.

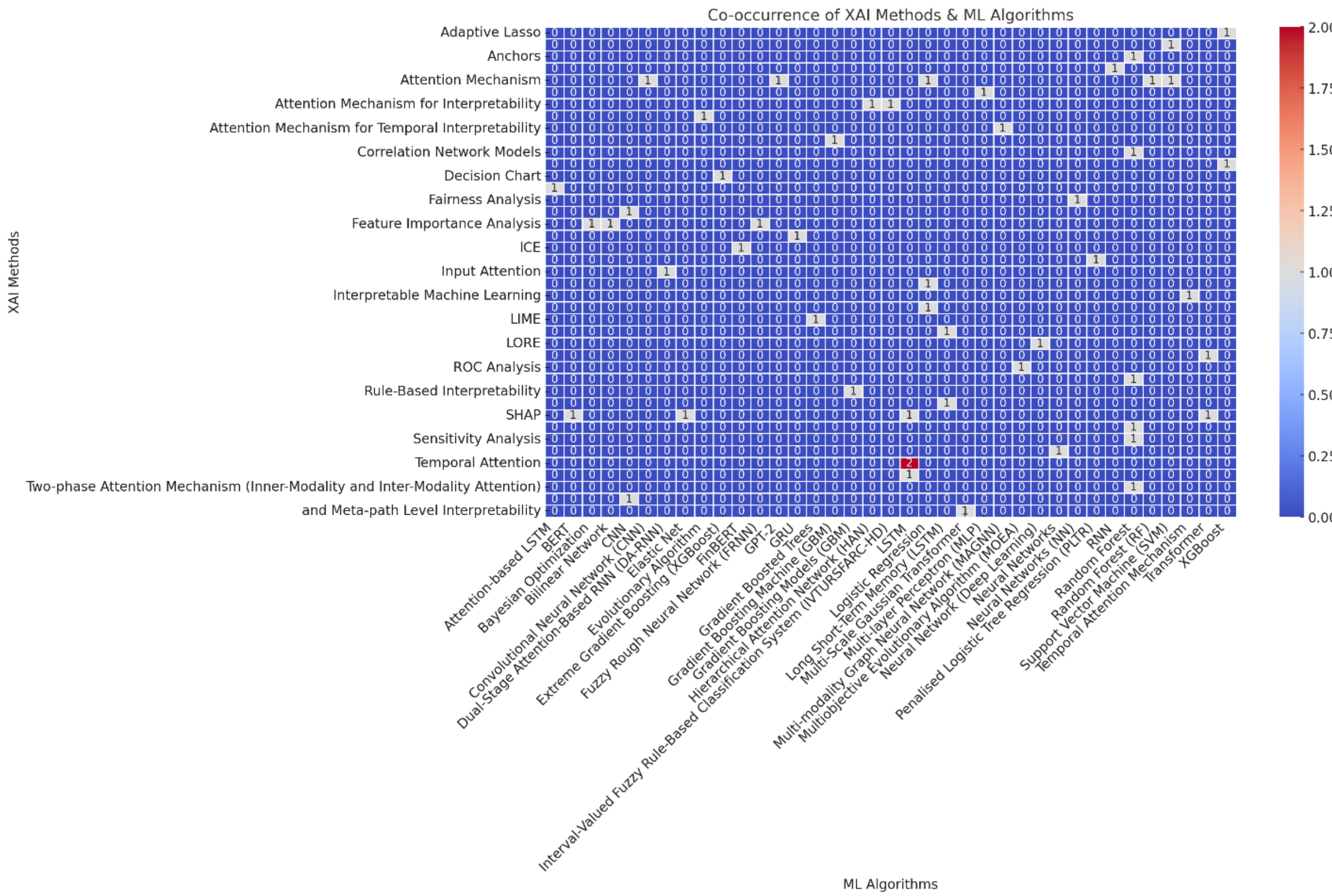

**Fig 11.** Co-occurrence Analysis of XAI Methods and ML Algorithms

Of the chosen studies, trade and investment category accounts for 26.5%, credit evaluation and scoring accounts for 23.5%, and fraud detection accounts for 20.6% of the study. The concentration on investment, risk reduction, and decision transparency shown by the high presence of these categories' points to their increasing importance and dominance. Many of these research use machine learning models—often combining explainability techniques as SHAP, LIME, and Feature Importance Scores to increase openness. Particularly in financial time series forecasting, stock price prediction, and portfolio optimization, a great deal of research center on trading and investment. To improve prediction accuracy researchers, use LSTM, Transformer-based architectures, and Reinforcement Learning models. In this field, attention methods are extensively applied to highlight the primary financial factors affecting market movements, hence enhancing model interpretability. Several publications also provide multi-objective optimization methods that balance explainability with forecast accuracy, therefore enabling financial analysts to better understand AI-driven trading strategies.

With many studies using network-based analysis, anomaly detection, and deep learning methods to identify fraudulent transactions, fraud detection is still a major focus of research. These researches seek to minimize false positives—a recurring difficulty in the financial industry—while simultaneously increasing fraud detection accuracy. Often used are approaches include hierarchical attention networks for deep learning-based fraud identification, random forests for transaction analysis, and graph-based fraud detection models. These models depend much on feature attribution and counterfactual explanations, which offer clear justifications for identified fraudulent activities. Comprising 14.7% of the chosen papers, research in risk management investigates the function of explainable artificial intelligence in credit risk assessment, financial regulation, and decision-making openness. Commonly utilized approaches include Shapley values, correlation network models, and counterfactual explanations. In banking and finance, where responsibility and justice in AI decision-making are vital, these studies show the increasing requirement of interpretable artificial intelligence models. The variety of methodological techniques used in different financial subfields emphasizes how rapidly hybrid machine learning and deep learning models are being accepted. Underlining the industry's need for open, interpretable, and regulation-compliant AI solutions, the most often employed explainability methodologies are SHAP, LIME, Attention Mechanisms, and Feature Attribution Scores.

Our analysis of the publication trends shows a notable rise in research post-2018 in line with the acceptance of explainability in finance. Extensively referenced are deep learning-based explainability methods for trading and investment, especially Transformer-based models and LSTM architecture used for portfolio optimization and stock price prediction. Many research proposes multi-head attention processes, which offer more thorough understanding of the main market factors affecting investment choices. These studies show how urgently interpretable artificial intelligence-driven trading techniques matching financial expert expectations are needed. Particularly studies using graph-based and anomaly detection algorithms have attracted a lot of references in fraud detection research. Many of these research center on financial transaction fraud detection using deep learning explainability methods to increase model openness. By means of saliency maps and hierarchical attention processes, financial analysts can grasp the reasons behind specific transaction flagging as fraudulent, thereby lowering false positives and raising detection accuracy. Research on risk management and regulatory compliance is also becoming more popular; papers on fairness in AI-driven financial systems are driving more citations for both. Research including causality-driven risk assessment systems, justice-aware artificial intelligence models, and counterfactual explanations shows a rising focus on guaranteeing ethical and fair financial AI models. Combining decision forests with correlation-adjusted models to improve interpretability while keeping good predictive accuracy, several articles provide ensemble solutions for reducing financial risk.

There are several limitations that are evident across the reviewed studies as well. (Qin et al. 2017) (Xia et al. 2017) (Moscato et al. 2021) suffers from limited real-time application. The need for further exploration of the model behavior is required in (Van Vlasselaer et al. 2015). Dataset constraints present another significant limitation in (Jin et al. 2020) (Dumitrescu et al. 2022) (Wang et al. 2022) (Chen and Ge 2019) (Zhong et al. 2020) (Florez-Lopez and Ramon-Jeronimo 2015). Significantly higher computational costs pose a constraint in studies such as (Cao et al. 2020) (Cheng et al. 2022) (Tran et al. 2018) (Huang et al. 2023) (Sanz et al. 2014) (Craja et al. 2020) (Antonelli et al. 2017) (Wang et al. 2019) (Van Vlasselaer et al. 2015) (Addo et al. 2018), where complex model architectures lead to significant processing overheads as well as limiting scalability.

### 4.3 Challenges in Implementing XAI in Financial domain

Though XAI holds great promise for finance, various obstacles have hampered its acceptance. AI systems should clearly explain their outputs; the way a system interacts with a user should never be taken as though the choice was made by a person rather than a machine (Purificato et al. 2023). However, developing models that are both accurate and interpretable is challenging due in great part to the complexity of financial data and the numerous links between variables. Moreover, even if XAI approaches may explain model decisions, there is usually a trade-off between model interpretability and accuracy since simpler models may offer more direct explanations but at the expense of lowered predictive performance. The ethical and legal ramifications of XAI in finance provide another difficulty especially in terms of guaranteeing justice, openness, and responsibility in artificial intelligence-driven decision-making. Last but not least, the absence of uniform evaluation systems for XAI in finance poses a major challenge since practitioners and academics lack common benchmarks to evaluate the success of several approaches. Many of the current artificial intelligence models in financial applications are still seen as "black-box" technology, which makes it challenging for stakeholders to completely trust or understand the decision-making process supporting AI projections. Among the several difficulties this lack of openness generates are ethical ones, questions about regulatory compliance, and doubts on the validity of decisions made by AI-driven companies. Therefore, resolving these challenges and supporting ethical AI deployment in the industry depends on investigating XAI approaches in the framework of finance.

# 5 Conclusion

In the last few years, AI has achieved a notable momentum that, if harnessed appropriately, may deliver the best of expectations in finance. Integration of XAI in finance reflects a basic movement toward more transparent, responsible, and interpretable financial decision-making. In this paper, the junction of XAI and finance has been methodically investigated, underlining important trends, methodological developments, and useful applications of explainability methodologies. Notwithstanding these developments, some issues still exist, especially in the scalability, real-time application, and XAI standardization in financial systems. As lack of interpretability and auditability of AI and Machine Learning (ML) methods could become a macro-level risk (Board 2017), it is imperative to learn how the inherent models work. Researchers, practitioners, and industry working together will help to create stronger, more scalable, and domain-specific explainability solutions from addressing these shortcomings. This improved interpretability and transparency would boost confidence in the models, allowing them to be widely used in real-world applications. As artificial intelligence (AI) continues to transform financial markets, the ability to explain and defend algorithmic models will be important for generating confidence, ensuring fairness, and meeting regulatory obligations. The rise of XAI in finance is more than just a technological advancement; it represents a necessary paradigm shift toward ethical, responsible, and transparent AI-powered financial systems.

# Appendix

**Table 2:** Taxonomy of XAI Applications

| Paper Name | Author | Publication Year | Category | Keywords | ML Algorithm | XAI Methods | Evaluation Metrics | Journal Name |
|---|---|---|---|---|---|---|---|---|
| A Dual-Stage Attention-Based Recurrent Neural Network for Time Series Prediction | (Qin et al. 2017) | 2017 | Time Series Forecasting | | Dual-Stage Attention-Based RNN (DA-RNN), LSTM, GRU | Input Attention, Temporal Attention | RMSE, MAE, MAPE | arXiv preprint |
| A Boosted Decision Tree Approach Using Bayesian Hyper-parameter Optimization for Credit Scoring | (Xia et al. 2017) | 2017 | Credit Evaluation and Scoring | Credit scoring, Boosted decision tree, Bayesian hyper-parameter optimization | Extreme Gradient Boosting (XGBoost), Bayesian Optimization | Feature Importance Scores, Decision Chart | Accuracy, Error Rate, AUC-H Measure, Brier Score | Expert Systems With Applications |
| APATE: A Novel Approach for Automated Credit Card Transaction Fraud Detection using Network-Based Extensions | (Van Vlasselaer et al. 2015) | 2015 | Fraud Detection | Credit card transaction fraud, network analysis, bipartite graphs, supervised learning | Logistic Regression, Neural Networks, Random Forest | Feature Importance Analysis | AUC, Accuracy, Balanced Accuracy | Decision Support Systems |
| Stock | (Jin, | 2020 | Trading | Stock market | LSTM, | Attention | MAE, RMSE, | Neural |
| closing price prediction based on sentiment analysis and LSTM | Yang, and Liu 2020) | | and Investment | prediction, Long short-term memory, Attention mechanism, Empirical mode decomposition | CNN | Mechanism | MAPE, $R^2$, Granger Causality | Computing and Applications |
| Explainable Machine Learning in Credit Risk Management | (Bussmann et al. 2021) | 2021 | Risk Management | Credit risk management, Explainable AI, Financial technologies, Similarity networks | XGBoost, Logistic Regression | Shapley Values, Correlation Network Models, TreeSHAP | AUC, Misclassification Rate, Receiver Operating Characteristics (ROC) | Computational Economics |
| Machine Learning for Credit Scoring: Improving Logistic Regression with Non-Linear Decision-Tree Effects | (Dumitrescu et al. 2022) | 2022 | Credit Evaluation and Scoring | Risk management, Credit scoring, Machine learning, Interpretability, Econo-metrics. | Penalised Logistic Tree Regression (PLTR), Random Forest, Support Vector Machine (SVM), Neural Networks (NN) | Feature Engineering with Short-Depth Decision Trees, Adaptive Lasso | AUC, Brier Score, Kolmogorov-Smirnov Statistic (KS), Percentage of Correct Classification (PCC), Partial Gini Index (PGI) | European Journal of Operational Research |
| Multiobjective Evolution of Fuzzy Rough Neural Network via Distributed Parallelism for Stock Prediction | (Cao et al. 2020) | 2020 | Trading and Investment | Distributed parallelism, evolutionary neural network, fuzzy rough neural network (FRNN), multiobjective evolution, stock price | Fuzzy Rough Neural Network (FRNN), Multiobjective Evolutionary Algorithm (MOEA) | Interval type-2 fuzzy neurons, IF-THEN rules-based interpretability | Prediction Precision (MAE, RMSE), Network Simplicity | IEEE Transactions on Fuzzy Systems |
| | | | | prediction | | | | |
| Financial time series forecasting with multi-modality graph neural network | (Cheng et al. 2022) | 2022 | Time Series Forecasting | Graph neural network, Graph attention, Deep learning, Quantitative investment | Multi-modality Graph Neural Network (MAGNN) | Two-phase Attention Mechanism (Inner-Modality and Inter-Modality Attention) | Micro-F1, Macro-F1, Weighted-F1, Accumulated Return, Daily Return, Sharpe Ratio | Pattern Recognition |
| Disparate Interactions: An Algorithm-in-the-Loop Analysis of Fairness in Risk Assessments | (Green and Chen 2019) | 2019 | Risk Management | Fairness, risk assessment, behavioral experiment, Mechanical Turk | Gradient Boosted Trees | Algorithm-in-the-Loop Framework, Fairness Analysis | Area Under Curve (AUC), False Positive Rate, Brier Score | FAT* '19 Conference on Fairness, Accountability, and Transparency |
| Credit Risk Analysis Using Machine and Deep Learning Models | (Addo, Guegan, and Hassani 2018) | 2018 | Risk Management | Credit risk, financial regulation, data science, Big Data, deep learning | Elastic Net, Random Forest, Gradient Boosting Machine (GBM), Neural Network (Deep Learning) | Feature Importance Analysis, ROC Analysis | AUC (Area Under Curve), Root Mean Square Error (RMSE), Gini Index | Risks |
| Temporal Attention-Augmented Bilinear Network for Financial Time-Series Data Analysis | (Tran et al. 2018) | 2018 | Time Series Forecasting | Bilinear projection, feedforward neural network, financial data analysis, temporal attention, time-series prediction. | Bilinear Network, Temporal Attention Mechanism | Attention Mechanism for Temporal Interpretability | Micro-F1, Macro-F1, Weighted-F1, Computational Complexity, Training Time | IEEE Transactions on Neural Networks and Learning Systems |
| A Benchmark | (Moscato , | 2021 | Credit Evaluati | Credit score prediction, | Logistic Regression, | LIME, SHAP, | AUC, Sensitivity | Expert Systems |
| of Machine Learning Approaches for Credit Score Prediction | Picariello, and Sperlí 2021) | | on and Scoring | Benchmark, Supervised learning, Machine learning, Explainable artificial intelligence | Random Forest, Multi-layer Perceptron (MLP) | Anchors, BEEF, LORE | (TPR), Specificity (TNR), Accuracy (ACC), G-Mean, False Positive Rate | with Applications |
| (Huang, Wang, and Yang 2023) | | 2023 | Financial Analysis and Decision-Making | Deep learning, large language model, transfer learning, interpretable machine learning, sentiment classification, environment, social, and governance (ESG) | BERT, FinBERT, Convolutional Neural Network (CNN), Long Short-Term Memory (LSTM), Support Vector Machine (SVM), Random Forest (RF) | Feature Importance Analysis, Interpretable Machine Learning | Accuracy, Precision, Recall, F1 Score, AUC | Contemporary Accounting Research |
| A Compact Evolutionary Interval-Valued Fuzzy Rule-Based Classification System for the Modeling and Prediction of Real-World Financial Applications With Imbalanced Data | (Sanz et al. 2014) | 2014 | Financial Modeling and Prediction | Evolutionary algorithms, financial applications, interval-valued fuzzy rule-based classification systems, intervalvalued fuzzy sets (IVFSs). | Interval-Valued Fuzzy Rule-Based Classification System (IVTURSFARC-HD), Evolutionary Algorithm | Interpretability through Fuzzy Rules, Rule Weight Rescaling Method | Geometric Mean (GM), True Positive Rate (TPR), True Negative Rate (TNR), Accuracy | IEEE Transactions on Fuzzy Systems |
| Hierarchical Multi-Scale Gaussian Transformer for Stock Movement Prediction | (Ding et al. 2020) | 2020 | Trading and Investment | Transformer, Multi-Scale Gaussian Prior, Orthogonal Regularization, Trading Gap Splitter, Stock Movement Prediction | Transformer, Multi-Scale Gaussian Transformer | Attention Mechanism for Hierarchical Interpretability | Accuracy, Matthews Correlation Coefficient (MCC) | International Joint Conference on Artificial Intelligence (IJCAI-20) |
| An Explainable AI Decision-Support System to Automate Loan Underwriting | (Sachan et al. 2020b) | 2020 | Credit Evaluation and Scoring | Explainable artificial intelligence, Interpretable machine learning, Loan underwriting, Evidential reasoning, Belief-rule-base Automated decision making | Gradient Boosting Models (GBM), Random Forest | SHAP, ICE | Loan Approval Rates, Model Fairness | Expert Systems with Applications |
| Deep Learning for Detecting Financial Statement Fraud | (Craja, Kim, and Lessmann 2020) | 2020 | Fraud Detection | Fraud detection, Financial statements, Deep learning, Text analytics | Hierarchical Attention Network (HAN), LSTM, Random Forest, XGBoost, GPT-2 | Attention Mechanism, LIME (Local Interpretable Model-Agnostic Explanations) | AUC, Sensitivity, Specificity, F1-Score, F2-Score, Accuracy | Decision Support Systems |
| Stock market index prediction using deep Transformer model | (Wang et al. 2022) | 2022 | Trading and Investment | Deep learning, Transformer, Stock index prediction | Transformer, CNN, RNN, LSTM | Attention Mechanism | Mean Absolute Error (MAE), Mean Squared Error (MSE), Mean Absolute Percentage Error (MAPE), Sharpe Ratio, | Expert Systems With Applications |
| | | | | | | | Volatility, Max Drawdown | |
| Exploring the Attention Mechanism in LSTM-based Hong Kong Stock Price Movement Prediction | (Chen and Ge 2019) | 2019 | Trading and Investment | LSTM, Stock price, Prediction | Long Short-Term Memory (LSTM), Attention-based LSTM | Attention Mechanism | Accuracy, Precision, Recall, F1 Score | Quantitative Finance |
| A Deep Learning Approach for Credit Scoring of Peer-to-Peer Lending Using Attention Mechanism LSTM | (Wang et al. 2018) | 2018 | Credit Evaluation and Scoring | P2P lending, credit scoring, machine learning, deep learning, LSTM, attention mechanism | LSTM, Bi-directional LSTM (BLSTM), Attention Mechanism | Attention Mechanism for Interpretability | ROC Curve, AUC (Area Under Curve), Kolmogorov-Smirnov (KS) Statistic | IEEE Access |
| Financial Defaulter Detection on Online Credit Payment via Multi-view Attributed Heterogeneous Information Network | (Zhong et al. 2020) | 2020 | Fraud Detection | Financial defaulter detection, Multi-view attributed heterogeneous information network, Meta-path encoder | Meta-path based Path Encoder, Neural Networks, Attention Mechanism, Softmax Classifier | Attention Mechanism for Node, Link, and Meta-path Level Interpretability | AUC (Area Under the Curve), Recall@Precision=0.1 | The Web Conference (WWW) 2020 |
| Multiobjective Evolutionary Optimization of Type-2 Fuzzy | (Antonelli et al. 2017) | 2017 | Financial Analysis and Decision-Making | Financial datasets, multiobjective evolutionary fuzzy systems, | Type-2 Fuzzy Rule-Based Classifier, Multiobjective Evolutiona | Rule-Based Interpretability, Scaled Dominance Approach | Accuracy, Complexity (Number of Rules), True Positive Rate (TPR), False Positive Rate | IEEE Transactions on Fuzzy Systems |
| Rule-Based Systems for Financial Data Classification | | | | type-2 fuzzy rule-based classifiers, unbalanced datasets. | ry Algorithm (MOEA), C4.5 Decision Tree, FURIA | | (FPR) | |
| Financial Defaulter Detection on Online Credit Payment via Multi-view Attributed Heterogeneous Information Network | (Zhong et al. 2020) | 2020 | Fraud Detection | Cross-country study, national culture, individualism, stock price crash risk | Meta-path-based Path Encoder, Attention Mechanism, Multi-view Learning | Attention Mechanism for Interpretability | AUC (Area Under Curve), Recall@Precision=0.1 | Proceedings of The Web Conference (WWW) |
| Explainable AI in Fintech Risk Management | (Bussmann et al. 2020) | 2020 | Risk Management | Credit risk management, explainable AI, financial technologies, peer to peer lending, logistic regression, predictive models | Logistic Regression, XGBoost | SHAP | Accuracy, ROC-AUC | Frontiers in Artificial Intelligence |
| AlphaStock: A Buying-Winners-and-Selling-Losers Investment Strategy using Interpretable Deep Reinforcement | (Wang et al. 2019) | 2019 | Trading and Investment | Quantitative Trading, Portfolio Management | Reinforcement Learning, Deep Attention Networks, Long Short-Term Memory (LSTM) | Sensitivity Analysis, Cross-Asset Attention Mechanism | Sharpe Ratio, Annualized Percentage Rate (APR), Maximum Drawdown (MDD), Calmar Ratio (CR), Downside Deviation Ratio (DDR) | KDD '19 - ACM SIGKDD Conference |
| Attention Networks | | | | | | | | |
| Forecasting Daily Stock Trend Using Multi-Filter Feature Selection and Deep Learning | (Haq et al. 2021) | 2021 | Trading and Investment | Stock trend prediction, Feature selection, Deep learning, Machine learning | Logistic Regression, Support Vector Machine (SVM), Random Forest, Deep Generative Models, Variational AutoEncoder (VAE), Attention Mechanism | Attention Mechanism | Prediction Accuracy, Classification Accuracy, Sensitivity, Specificity | Expert Systems with Applications |
| Enhancing accuracy and interpretability of ensemble strategies in credit risk assessment | (Florez-Lopez and Ramon-Jeronimo 2015) | 2015 | Risk Management | Ensemble strategies, Credit scoring, Decision forests, Diversity, Gradient boosting, Random forests | Correlated-Adjusted Decision Forest (CADF) | SHAP | Accuracy, RMSE, Precision | Annals of Operations Research |
| Transformer-based Attention Network for Stock Movement Prediction | (Zhang et al. 2022) | 2022 | Trading and Investment | Stock movement prediction, Deep learning, Transformer, Attention | Transformer, LSTM, Multihead Attention, Temporal Attention | Visualization of Attention Scores (Multihead and Temporal Attention Mechanisms) | Accuracy, Matthews Correlation Coefficient (MCC) | Expert Systems with Applications |
| Accurate Multivariate Stock Movement Prediction | (Yoo et al. 2021) | 2021 | Trading and Investment | Stock movement prediction, transformers, attention | Data-axis Transformer with Multi-Level | Attention Map, Temporal Attention | Accuracy, Matthews Correlation Coefficient (MCC), | ACM SIGKDD Conference on Knowledg |
| via Data-Axis Transformer with Multi-Level Contexts | | | | mechanism | Contexts (DTML) | | Investment Simulation (Annualized Return) | e Discovery and Data Mining |
| Enhancing accuracy and interpretability of ensemble strategies in credit risk assessment. A correlated-adjusted decision forest proposal | (Florez-Lopez and Ramon-Jeronimo 2015) | 2015 | Fraud Detection | Ensemble strategies, Credit scoring, Decision forests, Diversity, Gradient boosting, Random forests | Gradient Boosting, Random Forests, Decision Trees | Rule-based reasoning, Decision Trees for Interpretability | Accuracy Rate, Type I & II Error, AUC, Statistical Tests | Expert Systems with Applications |